\documentclass[showpacs,pra,aps,superscriptaddress,twocolumn,floatfix]{revtex4-1}
\usepackage{amsmath}
\usepackage{amssymb,mathrsfs}
\usepackage{graphicx}
\newcommand{\la}{\lambda}
\newcommand{\K}{\mathrm{K}}
\newcommand{\E}{\mathrm{E}}
\newcommand{\prt}{\partial}
\newfont{\ensmathquatorze}{msbm10 scaled 1400}
\newfont{\ensmathonze}{msbm10 scaled 1100}
\newfont{\ensmathdix}{msbm10}
\newfont{\ensmathneuf}{msbm10 scaled 833}
\newfont{\ensmathhuit}{msbm10 scaled 694}
\newfam\ensmathfam
\textfont\ensmathfam=\ensmathonze
\scriptfont\ensmathfam=\ensmathdix
\scriptscriptfont\ensmathfam=\ensmathhuit
\def\sss{\scriptscriptstyle}
\def\s2{\scriptstyle}

\begin{document}

\title{Generation of dispersive shock waves by the flow of a
Bose-Einstein condensate past a narrow obstacle}

\author{A. M. Kamchatnov}\affiliation{Institute of Spectroscopy,
  Russian Academy of Sciences, Troitsk, Moscow Region, 142190, Russia}
\author{N. Pavloff}\affiliation{Univ. Paris Sud, CNRS, Laboratoire de
  Physique Th\'eorique et Mod\`eles Statistiques, UMR8626, F-91405
  Orsay, France}

\begin{abstract}
  We study the flow of a one-dimensional Bose-Einstein
  condensate incident onto a narrow obstacle. We consider a
  configuration in which a dispersive shock is formed and propagates
  upstream away from the obstacle while the downstream flow reaches a
  supersonic velocity, generating a sonic horizon. Conditions
  for obtaining this regime are explicitly derived and the accuracy of our
analytical results is confirmed by numerical simulations.
\end{abstract}


\maketitle

\section{Introduction}

The flow of a quantum fluid past an obstacle can ge\-ne\-ra\-te
various excitations. In the classical example of liquid {\rm HeII}
these are phonons and rotons introduced by Landau \cite{landau} in his
theory of superfluidity. According to Landau, superfluidity is lost
when the flow velocity past an obstacle or through a capillary tube
exceeds the threshold of creation of Cherenkov-like ra\-dia\-tion of
linear waves (or, in other words, of quantum quasi-particles). In
actual experiments where {\rm HeII} flows through a narrow channel,
superfluidity is lost at velocities much lower than what
is predicted by the Landau criterium (see, e.g., \cite{Wilks}).  This
discrepancy between theory and experiment was explained by Feynman
\cite{feynman} as re\-sul\-ting from the nucleation of {\it nonlinear}
excitations: vortex rings generated by the flow past the obstacle.

The same phenomenology is observed in the flow of Bose-Einstein
condensates (BEC), either for trapped ultracold atomic vapors or
polaritons in semiconductor mi\-cro\-ca\-vi\-ties, with special
features linked to the specific dispersion relation of elementary
excitations in these systems: for instance the Landau critical
velocity is the velocity of sound $c$, and in two dimensions (2D) the
Cherenkov radiation forms an interference pattern located outside of
the Mach cone \cite{WMCA-99,Car-2006,GEGK-2007,GSK-2008}. These
specificities being taken into account, one observes phenomena similar
to those observed in liquid helium: in 2D or 3D flows past obstacles,
su\-per\-flui\-di\-ty is broken at velocity lower than $c$
\cite{Raman-1999,Amo-2009} because of the nucleation of vortices
\cite{FPR-92,Inou-2001,Nee-2010,Nar-2011,San-2011} or generation of
effectively stable oblique solitons \cite{EGK-2006,kp-2008,kk-2011}
(recently observed in experiments with polariton condensates
\cite{amo-2011,gro11}), and more complicated dispersive shock waves
(DSW) patterns \cite{ek-2006,el-2009,hi-2009}.

\begin{figure}
\includegraphics*[width=0.99\linewidth]{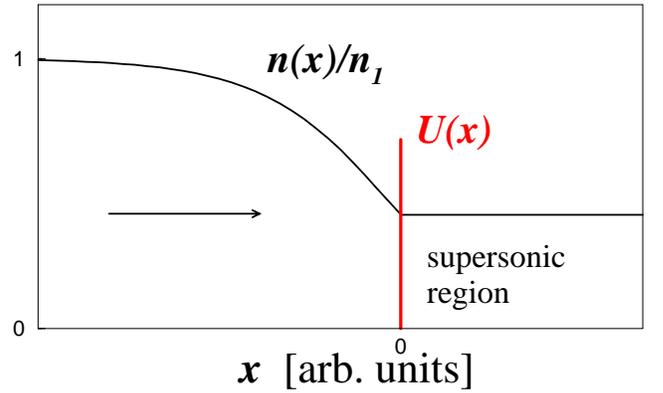}
\caption{(Color online) Density profile in vicinity of the  delta
  potential represented by a (red) vertical straight line.
The flow is stationary, with a velocity directed toward
  positive $x$ as indicated by the arrow. The density in the region
  $x<0$ is a portion of a dark soliton (see Sec. \ref{CetD}) with
  asymptotic upstream density ($n_1$) and subsonic velocity. The flow
  in the downstream region $x>0$ has a constant density and a supersonic
  velocity.}
\label{fig0}
\end{figure}

The situation in quasi-1D flows is similar: although generation of
vortices or oblique solitons is impossible in this case, dark solitons
are still easily generated stable nonlinear excitations
\cite{Bur-1999,Den-2000}.  Together with DSW, these nonlinear
excitations are, as is already the case in higher dimension, keys
ingredients for understanding the time-dependent dynamics of guided
BECs in presence of obstacles, as experimentally studied in
Refs.~\cite{Ono-2000,ea-2007,Dri-2010}.  Besides their interest for
studying non\-li\-near quantum transport, quasi-1D BEC flows have also
been suggested as model systems for creating sonic horizons suitable
for the experimental observation of acoustic analogues of Hawking
radiation
\cite{GACZ-2000,Gio-2004,BCGJ06,correlations,Mac09,RPC-2009,Lar-2011}. In
particular, an interesting type of stationary sonic horizon has been
identified in Refs.~\cite{Lar-2011,lp-2001}. It corres\-ponds to the
situation where an upstream subsonic region is separated from a
downstream supersonic one by an obstacle of the form of a delta
potential (see Fig.~\ref{fig0} and discussion below). The present work
is devoted to the study of the dynamics of the formation of this
con\-fi\-gu\-ra\-tion. We show that it can be obtained by launching a
BEC wave packet onto a localized obstacle (not necessarily a delta
peak \cite{Laval}). When the wave packet reaches the potential, the
density typically piles up in front of the obstacle, forming a plateau
accompanied by a DSW which is ejected upstream. We study the
characteristic features of this DSW in detail (both analytically and
numerically) and determine for which parameters (specific to the flow
and to the obstacle) the situation just described can be realized.

Theoretical analysis of 1D flows past an obstacle has already been
addressed in a number of papers (see, e.g.,
\cite{hakim,lp-2001,pavloff,radouani,legk}). In these papers, it was
found that there exist two critical velocities $u_{-}$ and $u_{+}$
($u_{-}<c<u_{+}$). The sub-critical flows whose velocity $u$ is below
$u_{-}$ are superfluid and generate no excitations in vicinity of the
obstacle; the super-critical ones with velocity above $u_+$ do not
generate nonlinear DSW but the Cherenkov radiation is effective and
there is no sonic horizon; at last, in the trans-critical region, for
flows whose velocity $u$ lies between the two critical values,
$u_{-}<u<u_{+}$, DSW are generated, generally speaking at both sides
of the obstacle, and only in a very narrow range of velocities both
DSW are detached from the obstacle. Hence, it seems difficult to reach a
situation such as the one illustrated in Fig.~\ref{fig0}.

Note however that the above quoted references
\cite{hakim,lp-2001,pavloff,radouani,legk} mainly con\-si\-de\-red
flows with identical asymptotic parameters at both sides far enough
from the obstacle. This choice of boun\-da\-ry conditions is natural
when the obstacle is put into motion starting from a situation where
it is immobile in a condensate at rest.  However, another setup is
possible, which is of considerable practical interest, being similar
to the flow of a fluid through a Laval nozzle \cite{LL-6}. In this
configuration (already considered in Ref. \cite{Dekel2010}) the
parameters of the flow are kept fixed only on the side of the incoming
flow and the downstream flow expands freely into vacuum. In this case
a DSW can be created only upstream and the parameters of the shock can
be calculated analytically for two typical models of obstacle
potential: (i) a smooth potential with typical size greater than the
characteristic healing length $\xi$ and (ii) a short-range potential
acting at distances much less than $\xi$. In the first case the
dispersion effects can be neglected at the location of the obstacle
(in the so called hydraulic appro\-xi\-ma\-tion) and the theory
reduces to the scheme presented in Ref.~\cite{legk}. In the second
case the potential can be approximated by a $\delta$ function and its
action amounts to the matching condition of exact solutions at the
point of its location. This last case has not been considered so far
in this kind of problem and we shall discuss it here in detail. We
will show that it makes it possible to realize asymptotically (i.e.,
at large time) a sonic horizon such as represented in Fig.~\ref{fig0}.

The paper is organized as follows. In Sec. \ref{secII} we present the
problem and the typical dynamical situation we aim at describing.  In
Sec. \ref{secIII} we discuss the time-dependent analytical solutions
of the flow in each of the characteristics regions of space identified
in the previous section. In Sec. \ref{secV} we briefly compare our
results with the ones obtained in the case of a thick obstacle.  In
Sec. \ref{secIV} we compare the analytical results with numerical
si\-mu\-la\-tions and propose an improvement of the analytical
description. Finally we present our conclusions in
Sec. \ref{discussion}.

\section{Formulation of the problem}\label{secII}

In the so-called 1D-mean field regime \cite{Men02} the dy\-na\-mics of the
condensate is described by the Gross-Pitaevskii equation which governs
the evolution of the wave function $\psi(x,t)$. Expressing densities
in units of a re\-fe\-ren\-ce density $n_{\rm ref}$, energies, distances and
velocities in units of the corresponding chemical potential $\mu_{\rm
  ref}(n_{\rm ref})$, healing length $\xi_{\rm ref}=\hbar/(m \mu_{\rm
  ref})^{1/2}$ and speed of sound $c_{\rm ref}=\hbar/(m \xi_{\rm ref})$, one
can write the Gross-Pitaevskii equation in the form
\begin{equation}\label{3-1}
    {\rm i} \, \psi_t =
- \tfrac{1}{2}\psi_{xx}+\left[U(x) + |\psi|^2\right]\psi
\; .
\end{equation}
The Gross-Pitaevskii equation is a sufficient ingredient for
describing the generation of DSW in a BEC as shown by the comparisons between
theory and experiments displayed in Refs. \cite{Hoe2006,Mepp2009}.

By means of the Madelung substitution
\begin{equation}\label{3-2}
    \psi(x,t)=\sqrt{n(x,t)}\exp\left({\rm i}
\int^x u(x',t)\,{\rm d}x'\right)\,{\rm e}^{-{\rm i}\mu t}
\end{equation}
the Gross-Pitaevskii equation can be cast into an hydrodynamic-like form
for the density $n(x,t)$ and the flow velocity $u(x,t)$:
\begin{equation}\label{3-3}
    \begin{split}
    n_t+(n u)_{x}=0\; ,\\
    u_t+uu_{x}+n_{x}+\left(\frac{n_{x}^2}{8n^2}
    -\frac{n_{xx}}{4n}\right)_{x}=-U_{x} \; .
    \end{split}
\end{equation}

We now briefly introduce the concept of Riemann invariant by
considering the very simple example of the dispersionless limit of
Eqs. (\ref{3-3}). This case is not treated only for pedagogical
purposes. At the boundary between the DSW and regions of flat profiles
(where dispersion is indeed negligible) it will make it possible to
match the description of the non linear wave in terms of elaborate
Riemann variables [Eqs. (\ref{4-2}) and (\ref{4-3})] with a simple
des\-crip\-tion of the type specified by Eqs. (\ref{3-3b}), (\ref{3-3c1})
and (\ref{3-3c2}) [this matching will be achieved by means of
Eqs. (\ref{6-2a}) and (\ref{5-2a})].

The last term of the left hand-side of the last of
Eqs.~(\ref{3-3}) is responsible for the dispersive character of the
BEC wave. In the absence of potential, and in a regime where the
effects of dispersion can be neglected, Eqs.~(\ref{3-3}) reduce to
\begin{equation}\label{3-3a}
n_t+(n u)_x=0 \; , \quad u_t +u u_x + n_x=0 \; .
\end{equation}

These equations can be written in a more symmetric form by introducing the
following Riemann invariants
\begin{equation}\label{3-3b}
\lambda_{\pm}(x,t)=\frac{u(x,t)}{2}\pm\sqrt{n(x,t)} \; ,
\end{equation}
which evolve according to the equations [equi\-va\-lent to
(\ref{3-3a})] :
\begin{equation}\label{3-3c1}
\partial_t\lambda_\pm +v_\pm(\lambda_+,\lambda_-)\,
\partial_x\lambda_\pm=0 \; ,
\end{equation}
with
\begin{equation}\label{3-3c2}
  v_\pm(\lambda_+,\lambda_-)
  =\tfrac{1}{2}(3\lambda_\pm+\lambda_\mp) \; .
\end{equation}
We will encounter below other Riemann invariants which describe the
DSW and obey equations similar to (\ref{3-3c1}) and (\ref{3-3c2})
[Eqs.~(\ref{4-7}) and (\ref{vi})]. However, the dispersive nature of
the shock will be there fully taken into account, in contrast to the
simple approximation (\ref{3-3a}), (\ref{3-3c1}) and (\ref{3-3c2}).

\begin{figure}[bt]
\begin{center}
\includegraphics*[width=0.99\linewidth]{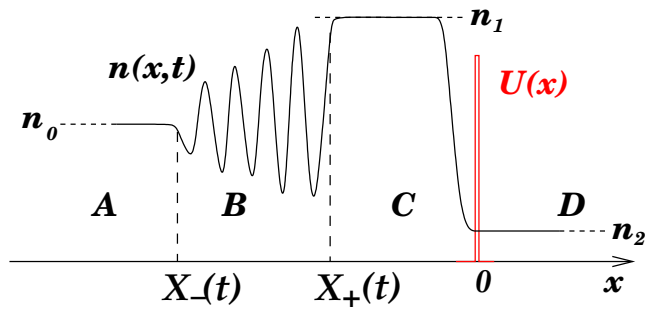}
\caption{(Color online) Sketch of the density profile $n(x,t)$ at a
  given time $t>0$. The flow is directed toward positive $x$.
The regions denoted as $A$, $B$, $C$ and $D$ are
  presented in the text. Points $X_-(t)$ and $X_+(t)$ are the
  small amplitude edge and soliton edge of the DSW (region $B$). The
  characteristic densities $n_0$, $n_1$ and $n_2$ are defined in the
  text [Eqs.~(\ref{3-51}), (\ref{5-2}) and (\ref{5-3})].}\label{fig1}
\end{center}
\end{figure}

Let us now present the initial velocity and density profile of the
incident flow. We suppose that at initial time ($t=0$) a half-infinite
pulse of BEC with a step-like distribution
\begin{equation}\label{3-8}
    \psi(x,t=0)=\left\{
    \begin{array}{lcl}
    \sqrt{n_0}\,\exp\{{\rm i}u_0 x\}&\mbox{for} & x<0\; ,\\
    0 &\mbox{for} & x>0\; .
    \end{array}
    \right.
\end{equation}
collides with a repulsive $\delta$ potential located at the origin of
the coordinate system:
\begin{equation}\label{3-7}
    U(x)=\kappa\, \delta(x)\; ,\quad \kappa>0 \; .
\end{equation}
In other words, the initial density and flow velocity are
\begin{equation}\label{3-51}
    n(x,t=0)=\left\{
    \begin{array}{lcl}
    n_0&\mbox{for} & x<0\; ,\\
    0 &\mbox{for} &  x>0\; ,
    \end{array}
    \right.
\end{equation}
\begin{equation}\label{3-52}
    u(x,t=0)=\left\{
    \begin{array}{lcl}
    u_0(>0) &\mbox{for} &x<0\; ,\\
    0 &\mbox{for} & x>0\; .
    \end{array}
    \right.
\end{equation}
At later times, the typical situation we aim at des\-cri\-bing is
illustrated in Fig.~\ref{fig1} which represents the density
distribution $n(x,t)$ at some fixed time $t>0$. It corresponds to a
flow initiated by the profile (\ref{3-51}) and (\ref{3-52}) in
which a plateau develops upstream of the potential (region $C$) while
a dispersive shock wave (region $B$) propagates away from the obstacle,
in the negative direction. The flow just downstream from the obstacle forms a
supersonic plateau (region $D$). The right edge of region $D$ matches
with a simple-wave solution (not shown in Fig. \ref{fig1}) which
des\-cri\-bes how the density vanishes at large $x$. In the present
work we are not interested in the description of this part of the flow
pattern. As one can see, the typical flow we consider can be
subdivided in this case into se\-ve\-ral regions (denoted as $A$, $B$, $C$
and $D$ in Fig.~\ref{fig1}) with specific features in each region.

$\bullet$ In region $A$ [$x<X_-(t)$] we have the incoming flow
$\psi_{\sss A}$ with the parameters (\ref{3-51}) and (\ref{3-52}). This flow
can be considered as a boundary condition:
\begin{equation}\label{4-1}
   \left\{\begin{array}{l} n_{\sss A}(x,t)=n_0\; ,\\ 
u_{\sss A}(x,t)=
u_0 \; ,\end{array}\right.
\quad x<X_- (t) \; , \quad\text{for all}\quad t>0.
\end{equation}
In this region there is no dispersion nor external potential and the
analysis in terms of Riemann invariants (\ref{3-3b}) is trivially
valid: the Riemann invariants are constant with $\la_\pm^{\sss
  A}=\frac12 u_0\pm\sqrt{n_0}$.

$\bullet$ In region $B$ [$X_{-}(t)<x<X_+(t)$] a dispersive shock wave
takes place which can be described as a modulated periodic solution
$\psi_{\sss B}(x,t)$ of Eqs.~(\ref{3-1}) and (\ref{3-3}) with
$U(x)=0$. Such a solution can be written in the form (see, e.g.,
\cite{kamch2000,legk})
\begin{equation}\label{4-2}
\begin{split}
    n_{\sss B}(x,t)= &\tfrac14(\la_4-\la_3-\la_2+\la_1)^2+\\
 &(\la_4-\la_3)(\la_2-\la_1)\times \\
     &\mathrm{sn}^2\left(\sqrt{(\la_4-\la_2)(\la_3-\la_1)}\,(x-Vt),m\right) \; ,
\end{split}
\end{equation}
\begin{equation}\label{4-3}
    u_{\sss B}(x,t)=V - \frac{C}{n_{\sss B}(x,t)}\; ,
\end{equation}
where $\mathrm{sn}$ is the sine elliptic Jacobi function,
\begin{equation}\label{4-41}
\begin{split}
   &V=\tfrac12\sum_{i=1}^4\la_i\; ,\quad
    m=\frac{(\la_2-\la_1)(\la_4-\la_3)}{(\la_4-\la_2)(\la_3-\la_1)}\; ,
\end{split}
\end{equation}
and
\begin{equation}\label{4-42}
\begin{split}
C=\tfrac18
&(-\la_1-\la_2+\la_3+\la_4)\times 
(-\la_1+\la_2-\la_3+\la_4)\times \\
    &(\la_1-\la_2-\la_3+\la_4) \; .\end{split}
\end{equation}

In strictly periodic solutions the parameters
$\la_1\leq\la_2\leq\la_2\leq\la_4$ are constant and they determine
characteristics of the wave such as
the amplitude of oscillations
\begin{equation}\label{4-5}
    a=(\la_2-\la_1)(\la_4-\la_3)\; ,
\end{equation}
and the wavelength
\begin{equation}\label{4-6}
L=\frac{2\,{\K}(m)}{\sqrt{(\la_4-\la_2)(\la_3-\la_1)}} \; ,
\end{equation}
$\K(m)$ being the complete elliptic integral of the first kind. In the
limit $m\to 0$ ($\lambda_2=\lambda_1$), (\ref{4-2}) describes a small
amplitude sinusoidal wave, and in the opposite case $m\to 1$
($\lambda_2=\lambda_3$) it describes a dark soliton.

In the case of a slowly modulated dispersive shock wave such as
occurring in region $B$, the
$\lambda$'s are functions of $x$ and $t$ which vary weakly over one
wavelength and one period. Their slow evolution is governed by the
Whitham equations \cite{whitham74,kamch2000}
\begin{equation}\label{4-7}
    \frac{\prt\la_i}{\prt t}+
v_i(\la_1,\la_2,\la_3,\la_4)\, \frac{\prt\la_i}{\prt x}=0,\quad i=1,2,3,4.
\end{equation}
Comparing with Eqs.~(\ref{3-3c1}) and (\ref{3-3c2}) one sees that the
$\la_i$'s are the Riemann invariants of the Whitham equations.  The
$v_i$'s are the characteristic velocities. Their $\lambda_i$'s
dependence is much more complicate than the simple linear combinations
appearing in (\ref{3-3c2}); they can be expressed in terms of complete
elliptic integrals of the first and the second kind
\cite{fl86,pavlov87}. One can use the following convenient formula for
their computation \cite{gke92,kamch2000}
\begin{equation}\label{vi}
v_i= V-\frac 12 \frac{L}{\partial L/\partial \la_i} ,\quad
i=1,2,3,4\; .
\end{equation}
At $x=X_-(t)$ we have the ``small amplitude edge'' of the dispersive shock
wave [with $\la_2(X_-(t),t)=\la_1(X_-(t),t)$] where the wave should
satisfy the matching conditions with the flow in the region $A$. This implies
that the mean values of the density $\overline{n}_{\sss B}$ and the flow
velocity $\overline{u}_{\sss B}$ coincide with $n_0$ and $u_0$
respectively:
\begin{equation}\label{5-1}
  \overline{n}_{\sss B}(X_-(t),t)=n_0,\quad
\overline{u}_{\sss B}(X_-(t),t)=u_0.
\end{equation}
Since at the left edge of the DSW we have $\la_2=\la_1$ the Whitham
equations (\ref{4-7}) for $\la_3$ and $\la_4$ can be shown to simplify to
\begin{equation}\label{6-2}
\begin{split}
\prt_t \la_3+& \tfrac{1}{2}\left(3\la_{3} + \la_{4}\right)
    \prt_x \la_3=0 \; , \\
\prt_t \la_4+& \tfrac{1}{2} \left(\la_{3}
+ 3\la_{4}\right)\prt_x \la_4=0 \; ,
\end{split}
\end{equation}
and these equations can be identified with the Riemann form
(\ref{3-3c1}) and (\ref{3-3c2}) of the dispersionless limit (\ref{3-3a}) of
Eqs.~(\ref{3-3}) (without potential); i.e., one has
\begin{equation}\label{6-2a}
\begin{split}
& \lambda_4(X_-(t),t)=\lambda^{\sss A}_+=\tfrac{1}{2}u_0+\sqrt{n_0}\equiv
\lambda_0 \; , \\
& \la_3(X_-(t),t)=\la^{\sss A}_-=\tfrac{1}{2}u_0-\sqrt{n_0}\; .
\end{split}
\end{equation}
The other edge of the DSW occurs at $x=X_+(t)$. We denote it as the
``soliton edge'' because at this point the density oscillations are
soliton-like : $\la_2(X_+(t),t)=\la_3(X_+(t),t)$ (i.e., $m=1$). The
matching conditions at $X_+$ read
\begin{equation}\label{5-2}
    \overline{n}_{\sss B}(X_+(t),t)=n_1\;
,\quad \overline{u}_{\sss B}(X_+(t),t)=u_1 \; ,
\end{equation}
where we suppose that $X_+$ is located far enough from the origin in order 
that the stationary solution $\psi_{\sss C}$ in region $B$
reaches its asymptotic values $n_{\sss C}(x)\to n_1$,
$u_{\sss C}(x)\to u_1$ when $x\simeq X_+$ 
(i.e., in the formal limit $x\to-\infty$ in region $B$, see below
for more details).

Since at the soliton edge $\lambda_2=\lambda_3$, one can show as
previously that the Whitham equations for $\la_4$ and $\la_1$ reduce
to a form similar to (\ref{3-3c1}) and (\ref{3-3c2}) where $\la_4$
plays the role of $\la_+$ and $\la_1$ plays the role of $\la_-$. Hence
one has
\begin{equation}\label{5-2a}
\begin{split}
& \lambda_4(X_+(t),t)=\lambda^{\sss C}_+=\tfrac{1}{2}u_1+\sqrt{n_1} \; , \\
& \la_1(X_+(t),t)=\la^{\sss C}_-=\tfrac{1}{2}u_1-\sqrt{n_1}\; .
\end{split}
\end{equation}

$\bullet$ The region $C$ corresponds to a smooth and sta\-tio\-na\-ry
solution $\psi_{\sss C}(x,t)$ of the Gross-Pitaevskii equation for
$X_+<x<0$ and the region $D$ to a smooth and stationary solution
$\psi_{\sss D}$ for $0<x<X_{\rm sw}$.  The flow at $x>X_{\rm sw}$
represents a simple wave solution (with small dispersive
corrections) connecting region $D$ with vacuum and is of no
interest to us (this is the reason why we do not show the region
$x>X_{\rm sw}$ in Fig.~\ref{fig1}). In region $D$ we have a uniform
flow
\begin{equation}\label{5-3}
\left\{\begin{array}{l}    
n_{\sss D}(x,t)=n_2=\mathrm{const}\; ,\\ 
u_{\sss D}(x,t)=u_2=\mathrm{const}\; ,\end{array}\right.
\quad \text{for}\quad    0<x<X_{\rm sw}.
\end{equation}
Since the flow is stationary in both regions $C$ and $D$, the
parameters (\ref{5-2}) and (\ref{5-3}) must satisfy the condition of
conservation of flux
\begin{equation}\label{5-4}
    n_1u_1=n_2u_2 \; ,
\end{equation}
and the Bernoulli law (constant value of the chemical potential $\mu$)
\begin{equation}\label{5-5}
    \tfrac12u_1^2+n_1=\tfrac12u_2^2+n_2=\mu \; .
\end{equation}
One must also satisfy the condition of continuity of the
wave function at $x=0$,
\begin{equation}\label{5-6}
    \psi_{\sss C}(0,t)=\psi_{\sss D}(0,t) \; ,
\end{equation}
and the jump condition for the derivative of the wave function which
follows from Eqs.~(\ref{3-1}) and (\ref{3-7})
\begin{equation}\label{5-7}
    \partial_x \psi_{\sss D}(0,t)-
\partial_x\psi_{\sss C}(0,t)=2\,\kappa\,\psi_{\sss C}(0,t).
\end{equation}
Thus, our task is to find the solution which satisfies all the above
matching conditions [Eqs.~(\ref{5-1}), (\ref{5-2}), (\ref{5-4}), (\ref{5-5}),
(\ref{5-6}) and (\ref{5-7})] and yields the parameters $n_1,\,u_1,\,n_2,\,u_2$
and the positions $X_{\pm}(t)$ of the edges of the DSW as functions of
the incoming flow parameters $n_0,\,u_0$ and of the potential strength
$\kappa$.

It is clear that such a solution does not exist for any choice of the
parameters $n_0,\,u_0,\,\kappa$. For example, in the limit $\kappa\to
0$ our system reduces to the well-known ``dam problem''
\cite{whitham74,kamch2000} which is described in the hydrodynamic approximation
by a simple wave solution without formation of dispersive shock wave.
Therefore our solution should be complemented by an explicit statement
of its conditions of existence.

\section{Solution of the problem}\label{secIII}

\subsection{Dispersive shock wave (region $B$)}

As was indicated above, the DSW is described by four parameters
$\la_i$, $i=1,\,2,\,3,\,4,$ which change slowly along the wave. We
suppose that the DSW is detached from the obstacle and propagates upstream
with the velocities $V_{\pm}$ of the edge points $X_{\pm}$. The distance
at which $\psi_{\sss C}$ reaches its asymptotic value (with $n_{\sss
  C}\simeq n_1$, $u_{\sss C}\simeq u_1$) when one goes away from the origin
is of order of the healing
length in the region $C$, i.e. $\sim n_1^{-1/2}$. Since this distance
is much less than the (time increasing) length $|X_+|$, we can assume in
good approximation that the formation of the plateau $n_{\sss C}\simeq
n_1$, $u_{\sss C}\simeq u_1$ occurs ins\-tan\-taneously.  Then the DSW can be
described by a self-similar solution of the Whitham equations
(\ref{4-7}) and for the left-propagating DSW we get
\begin{equation}\label{6-1}
    \la_1=\mathrm{const}\; ,
\quad \la_3=\mathrm{const}\; ,\quad
    \la_4=\mathrm{const}\; ,
\end{equation}
and
\begin{equation}\label{6-1a}
v_2(\la_1,\la_2(x/t),\la_3,\la_4)=\frac{x}t\; .
\end{equation}
This is the so-called Gurevich-Pitaevskii problem introduced into the
theory of DSW in Ref.~\cite{gp74} and used for description of internal
waves generated by the flow of water past an uneven bottom in
\cite{gs-1986,smyth-1987} (see also \cite{legk} and references
therein).

\begin{figure}[bt]
\begin{center}
\includegraphics*[width=0.99\linewidth]{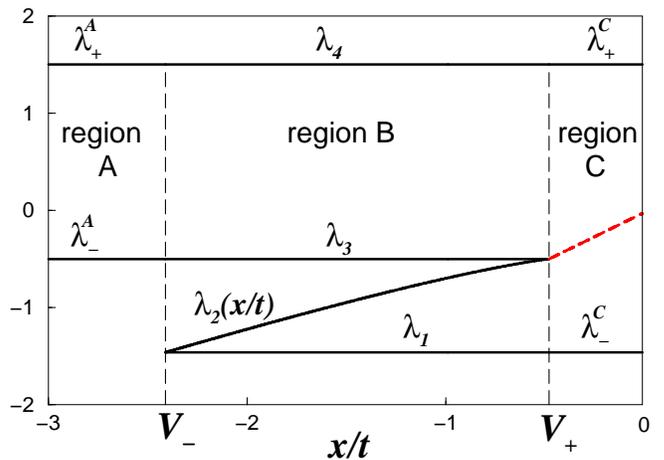}
\caption{(Color online) The Riemann invariants plotted as functions of
  the self-similar variable $x/t$ in regions $A$, $B$ and $C$. In
  region $C$, the (red) dashed line (not considered in Section
  \ref{secIII}) corresponds to the soliton train with
  $\la_2=\la_3=x/t-u_1/2$ discussed in Section \ref{secIV}
  [Eq.~(\ref{17-1})]. The figure is drawn for the situation studied in
  Sec. \ref{secIV}, with $n_0=1$, $u_0=1$ and $\kappa=5.2$. The corresponding
  velocities of the edges of the DSW (region $B$) are $V_-=-2.4$ and
  $V_+=-0.48$.}
\label{fig2}\end{center}
\end{figure}

When $n_1$ and $u_1$ are found, the values of the constants $\lambda_1$,
$\lambda_3$ and $\lambda_4$ can be determined from Eqs.~(\ref{6-2a})
and (\ref{5-2a}) whereas Eqs.~(\ref{vi}) and (\ref{6-1a}) determine
$\la_2$ as a function of $x/t$:
\begin{equation}\label{7-4}
\begin{split}
    & v_2(\la_1,\la_2,\la_3,\la_4)=\\
& \tfrac12\sum_i\la_i+\frac{(\la_3-\la_2)(\la_2-\la_1)\K(m)}
    {(\la_3-\la_2)\K(m)-(\la_3-\la_1)\E(m)}=
\frac{x}t\; .
\end{split}
\end{equation}
The plots of $\la_i$, $i=1,2,3,4$, as functions of $x/t$ are displayed in
Fig.~\ref{fig2}.
The left edge of the DSW moves with velocity
\begin{equation}\label{7-5}
\begin{split}
   V_-& \equiv\frac{X_-}t=v_2(\la_1,\la_1,\la_3,\la_4)\\
& = \tfrac12(2\la_1+\la_3+\la_4)+
    \frac{2(\la_3-\la_1)(\la_4-\la_1)}{2\la_1-\la_3-\la_4} \; ,
\end{split}
\end{equation}
and the right edge moves with velocity
\begin{equation}\label{7-6}
    V_+\equiv\frac{X_+}t=v_2(\la_1,\la_3,\la_3,\la_4)=
\tfrac12(\la_1+2\la_3+\la_4).
\end{equation}
For finding $n_1,\,u_1$ and,
hence, $\la_1$, we have to turn to the solution of the Gross-Pitaevskii
equation in region $C$ and its matching with the solution in region $D$.

\subsection{Flow across the $\delta$ potential (regions $C$ and
  $D$)}\label{CetD}

In regions $C$ and $D$ the flow is stationary, $\psi(x,t)=
{\rm e}^{-{\rm i}\mu t}\varphi(x)$,
so that the Gross-Pitaevskii equation reduces to
\begin{equation}\label{7-7}
    -\tfrac12\varphi_{xx}+|\varphi|^2\varphi+\kappa\delta(x)\varphi=
    \left(\tfrac12u_1^2+n_1\right)\varphi.
\end{equation}
It is convenient to introduce temporarily the variables
\begin{equation}\label{7-8}
\begin{split}
    \varphi=\sqrt{n_1}\phi,\quad u_1=\sqrt{n_1}\,M_{\scriptstyle c},\\ 
y=\sqrt{n_1}\,x,\quad
    \widetilde{\kappa}=\frac{\kappa}{\sqrt{n_1}},
\end{split}
\end{equation}
$M_{\s2 c}$ being the asymptotic (formally at $x\to-\infty$) Mach number of
the flow in region $C$. Then Eq.~(\ref{7-7}) takes the form
\begin{equation}\label{8-1}
    -\tfrac12\phi_{yy}+|\phi|^2\phi+\widetilde{\kappa}\delta(y)\phi=
    \left(\tfrac12M_{\s2 c}^2+1\right)\phi.
\end{equation}
From conservation of the flow (\ref{5-4}) we get
\begin{equation}\label{8-2}
    \frac{u_2}{u_1}=\frac{n_1}{n_2}\equiv\eta,
\end{equation}
and $\eta$ can be found as a function of $M_{\s2 c}=u_1/\sqrt{n_1}<1$ with the
use of Eq.~(\ref{5-5}) or
\begin{equation}\label{8-3}
    \frac{M_{\s2 c}^2}2\eta^3-\left(1+\frac{M_{\s2 c}^2}2\right)\eta+1=0.
\end{equation}
The relevant solution ($\eta>1$) of this equation is given by
\begin{equation}\label{8-4}
    \eta=\frac12\left(\sqrt{1+\frac8{M_{\s2 c}^2}}-1\right)
\; .
\end{equation}

We look for a solution of Eq.~(\ref{8-1}) of the form
\begin{equation}\label{8-5a}
    \phi(y)=
    e^{{\rm i}y M_{\s2 c}}\left\{\cos\theta\tanh[(y-y_0)\cos\theta]-{\rm i}
\sin\theta\right\},
\end{equation}
for $y<0$ and
\begin{equation}\label{8-5b}
    \phi(y)=-\frac1{\eta}e^{{\rm i}(y \eta M_{\s2 c}+\gamma)}\; ,
\end{equation}
for $y>0$. In the above expressions (\ref{8-5a}) and (\ref{8-5b}) we
have $u_2/\sqrt{n_1}=\eta M_{\s2 c}$ and $\sin\theta=M_{\s2 c}$.
This solution has been first identified in Ref. \cite{lp-2001}. 
The flow upstream from the obstacle corresponds to a portion of a dark
soliton which is attached at $y=0$ to a downstream supersonic flow.
The condition of continuity of the function $\phi(y)$ at $y=0$ [see
(\ref{5-6})] gives
\begin{equation}\label{8-6}
    \sin\gamma=\sqrt{\eta}\,\sin\theta,\quad 
\cos\gamma=\sqrt{\eta}\,\tanh(y_0\cos\theta) ,
\end{equation}
from which we get
\begin{equation}\label{8-7}
    \tanh(y_0\cos\theta)=
\sqrt{\frac{1-\eta M_{\s2 c}^2}{\eta(1-M_{\scriptstyle c}^2)}}.
\end{equation}
The condition (\ref{5-7}) takes the form
\begin{equation}\label{8-8}
    \phi_y(0^+)-\phi_y(0^-)=2\widetilde{\kappa}\phi(0) \; .
\end{equation}
Substitution of expressions (\ref{8-5a}) and (\ref{8-5b}) in this relation
gives after some algebra
with the use of Eqs.~(\ref{8-6}) and (\ref{8-7}) the equation
\begin{equation}\label{8-9}
    (\eta-1)\sqrt{\frac1{\eta^2}-M_{\s2 c}^2}=2\widetilde{\kappa}.
\end{equation}
Elimination of $\eta$ with the help of Eqs.~(\ref{8-4}) yields
\begin{equation}\label{8-10}
\widetilde{\kappa}=F(M_{\s2 c})
\quad\mbox{i.e.,}\quad
\kappa=\sqrt{n_1}\, F(M_{\s2 c}) \; ,
\end{equation}
where
\begin{equation}\label{8-11}
    F(M_{\s2 c})=\frac{M_{\s2 c}}8\left(\sqrt{1+\frac8{M_{\s2 c}^2}}-3\right)^{3/2}.
\end{equation}
Hence, according to (\ref{7-8})
\begin{equation}\label{9-2}
    u_1=\sqrt{n_1}\,M_{\s2 c}=\frac{\kappa M_{\s2 c}}{F(M_{\s2 c})}.
\end{equation}
Then Eqs.~(\ref{8-2}) and (\ref{8-4}) yield the expressions for $n_2$
and $u_2$:
\begin{equation}\label{9-3a}
    n_2=\frac{n_1}{\eta}=
\frac{\kappa^2M_{\s2 c}^2}{4F^2(M_{\s2 c})}
\left(\sqrt{1+\frac8{M_{\s2 c}^2}}+1\right),\
\end{equation}
\begin{equation}\label{9-3b}
    u_2=u_1\eta
=\frac{\kappa M_{\s2 c}}{2F(M_{\s2 c})}\left(\sqrt{1+\frac8{M_{\s2 c}^2}}-1\right).
\end{equation}
We thus have obtained the parameters $n_1,\,u_1,\,n_2,\,u_2$ as
functions of $\kappa$ and $M_{\s2 c}$. We now have to relate $M_{\s2
  c}$ to the physical parameters $n_0$ and $u_0$ which describe the
incoming flow.

\subsection{The global solution}

To get the global solution, we use the relation
$\la_+^A=\la_4=\la_+^{C}$ [see Eqs. (\ref{6-2a}) and (\ref{5-2a}) and
Fig.~\ref{fig2}] and (\ref{8-10}) and (\ref{9-2}) to obtain
\begin{equation}\label{9-4}
    \la_0\equiv\tfrac12u_0+\sqrt{n_0}=
\tfrac12u_1+\sqrt{n_1}=\frac{\kappa}{F(M_{\s2 c})}\left(1+\frac{M_{\s2 c}}2\right).
\end{equation}
As a result, all expressions can be written in a parametric form with
$M_{\s2 c}$ playing the role of the parameter. From (\ref{9-4}) we get
\begin{equation}\label{9-5}
    \kappa(\la_0,M_{\s2 c})=\la_0\frac{F(M_{\s2 c})}{1+M_{\s2 c}/2} \; ,
\end{equation}
which determines $M_{\s2 c}$ as a function of $\kappa$ and $\la_0$.
Substitution of this expression into (\ref{8-10})-(\ref{9-2}) yields
\begin{equation}\label{9-6}
    n_1(\la_0,M_{\s2 c})=
\frac{\la_0^2}{(1+M_{\s2 c}/2)^2},\; 
u_1(\la_0,M_{\s2 c})=\frac{\la_0M_{\s2 c}}{1+M_{\s2 c}/2},
\end{equation}
\begin{equation}\label{9-7a}
    n_2(\la_0,M_{\s2 c})=
\frac{\la_0^2M_{\s2 c}^2}{(2+M_{\s2 c})^2}
\left(\sqrt{1+\frac8{M_{\s2 c}^2}}-1\right),
\end{equation}
\begin{equation}\label{9-7b}
    u_2(\la_0,M_{\s2 c})=\frac{\la_0M_{\s2 c}}{2+M_{\s2 c}}
\left(\sqrt{1+\frac8{M_{\s2 c}^2}}+1\right).
\end{equation}
In combination with (\ref{9-5}), formulae (\ref{9-6}), (\ref{9-7a})
and (\ref{9-7b}) determine -- in a parametric form -- the dependence
of $n_1,\,u_1,\,n_2,\,u_2$ on $\kappa$ and
$\la_0=\frac{1}{2}u_0+\sqrt{n_0}$. These formulae represent an
important step in our study since they determine the overall structure
of the flow ($n_1,\,u_1,\,n_2$ and $\,u_2$) as a function of the initial
parameters
characterizing the incident beam ($n_0$ and $u_0$) and the obstacle
($\kappa$) with which it collides.

Actually, the overall structure of the flow is fully cha\-rac\-te\-ri\-zed
only once the locations $X_-(t)$ and $X_+(t)$ of the boundaries
between the different zones are known (cf. Fig. \ref{fig1}).  According
to our self-similarity hypothesis $X_{\pm}(t)=V_{\pm} \, t$, where
$V_{\pm}$ are the time-independent ve\-lo\-ci\-ties of the boundaries. Now
that $n_1$ and $u_1$ are found, $V_+$ and $V_-$ are simply determined
as follows. One first can get [from Eqs. (\ref{6-2a}) and (\ref{5-2a})]
the expressions for the 3 Riemann invariants which remain constant in
the region of the DSW
\begin{equation}\label{9-8}
\begin{split}
    \la_1=-\la_0\frac{2-M_{\s2 c}}{2+M_{\s2 c}},\quad 
\la_3=\tfrac12u_0-\sqrt{n_0},\\
    \la_4=\la_0=\tfrac12u_0+\sqrt{n_0} \; .
\end{split}
\end{equation}
Then, substitution of formulae (\ref{9-8}) into Eqs.~(\ref{7-5}) and
(\ref{7-6}) yields the desired expressions of the velocities $V_-$ and
$V_+$ of the edge points of the shock wave. In particular, it is
interesting to realize [cf. Eq.~(\ref{a4})] that the velocity
$V_-$ of the small amplitude edge $X_-$ of the DSW corresponds to the
group velocity of linear excitations in region $A$ of the system with
wavelength $L(X_-)$, where $L$ is the (position dependent) wavelength
of the nonlinear oscillations (\ref{4-2}). This is obtained by
noticing that the explicit formula (\ref{7-5}) and the expressions
(\ref{9-8}) of the Riemann invariants makes it possible to write $V_-$
under the form
\begin{equation}\label{a2}
\begin{split}
V_-=u_0+&2\sqrt{n_0}
\,\frac{1+M_{\s2 c}/2}{M_{\s2 c}-2-2M_0}\times\\
&\left[1+2 \frac{(2+M_0)(M_0-M_{\s2 c})}{(1+M_{\s2 c}/2)^2}\right] \; ,
\end{split}
\end{equation}
where $M_0=u_0/\sqrt{n_0}$.
Similarly, when $\la_1=\la_2$, the expression (\ref{4-6}) for $L$ reduces to
\begin{equation}\label{a3}
\begin{split}
L(X_-)=&\frac{\pi}{\sqrt{(\la_4-\la_1)(\la_3-\la_1)}}\\
=&\frac{\pi}{\sqrt{n_0}} 
\frac{(1+M_{\s2 c}/2)}{\sqrt{(M_0+2)(M_0-M_{\s2 c})}}\; .
\end{split}
\end{equation}
Defining $k=2\pi\sqrt{n_0}/L(X_-)$ as the dimensionless wave vector at
the small amplitude edge of the DSW, it is a simple exercise to show
that using Eq.~(\ref{a3}) one can rewrite (\ref{a2}) under the form
\begin{equation}\label{a4}
V_-=u_0-c_0 \frac{1+k^2/2}{\sqrt{1+k^2/4}}\; ,
\end{equation}
which is the group velocity of excitations whose dispersion relation
is given by $\omega(k)= u_0 k -c_0 k \sqrt{1+k^2/4}$, where
$c_0=\sqrt{n_0}$ is the speed of sound in a uniform system at rest
with density $n_0$. This dispersion relation corresponds to (left
propagating) linear elementary excitations in a BEC with constant
density $n_0$ and constant velocity $u_0$. Relation (\ref{a4}) is very
natural since $V_-$ is the velocity of the small amplitude edge $X_-$
at which the DSW is linear: the linear front of a wave-packet should
propagate with the appropriate group velocity. This is a generic
feature of the small amplitude edge of a DSW, as proven on general
grounds in Ref. \cite{El-saeDSW}.  Relation (\ref{a4}) nonetheless
constitutes a non-trivial check of the validity of our description,
since the connection between $V_-$ and $L(X_-)$ is not
straightforward.

As for the soliton edge of the DSW, we obtain from Eqs.~(\ref{7-6})
and (\ref{9-8})
\begin{equation}\label{10-1}
    V_+=\frac{u_0(1+M_{\s2 c})-2\sqrt{n_0}}{2+M_{\s2 c}}.
\end{equation}
This is the velocity of the edge soliton whose amplitude is equal to
\begin{equation}\label{10-2}
    a=(\la_3-\la_1)(\la_4-\la_3)=
\frac{4\sqrt{n_0}(u_0-M_{\s2 c}\sqrt{n_0})}{2+M_{\s2 c}}.
\end{equation}

Note that our approach not only determines the gross features of the
flow (those which are depicted in Fig. \ref{fig1}: $u_1$, $n_1$,
$u_2$, $n_2$ and $X_\pm$) but also yields a precise prediction for the
density profile in the region of the obstacle and of the DSW.  In the
region of the obstacle the order parameter is determined by
Eqs. (\ref{8-5a}) and (\ref{8-5b}). In the region of the DSW, the
velocity and the density profile are determined by Eqs. (\ref{4-2})
and (\ref{4-3}) which need the input of the $\la_i$'s [Eqs. (\ref{6-1a})
and (\ref{9-8})]. These predictions will be tested against numerical
simulations in Sec. \ref{secIV}.

\begin{figure}[bt]
\begin{center}
\includegraphics*[width=0.99\linewidth]{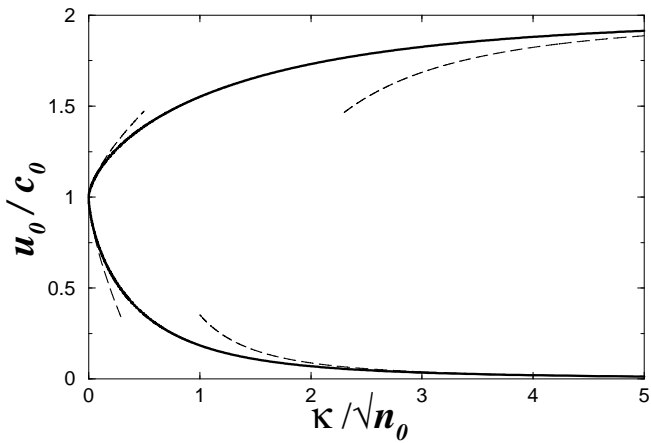}
\caption{Boundaries of the region of parameters for which the DSW is
  expelled upstream from the obstacle. The vertical axis is the speed of
  the incident beam in units of the speed of sound
  $c_0=\sqrt{n_0}$. The horizontal axis is the dimensionless strength
  $\kappa/\sqrt{n_0}$ of the $\delta$ potential. The DSW is expelled from the
  region of the obstacle when the point representative of the system
  lies between the two solid curves which correspond to
  Eqs.~(\ref{10-5}) and (\ref{10-6}). The dashed curves are the
  approximate boundaries (\ref{10-7}), (\ref{10-8}) (valid when
  $\kappa/\sqrt{n_0}\ll 1$), (\ref{11-1}) and (\ref{11-2}) (valid when
  $\kappa/\sqrt{n_0}\gg 1$).}
\label{fig3}\end{center}
\end{figure}

The solution found here is based on two assumptions: (i) that the
discontinuity $n_1>n_0$ arises after collision of the BEC pulse with
the obstacle. This is equivalent to the condition $a>0$ or
\begin{equation}\label{10-3}
    \frac{u_0}{\sqrt{n_0}}>M_{\s2 c} \; ;
\end{equation}
and (ii) that the DSW is detached from the obstacle, i.e., $V_+<0$ or
\begin{equation}\label{10-4}
    \frac{u_0}{\sqrt{n_0}}<\frac2{1+M_{\s2 c}}.
\end{equation}
If the density $n_0$ is fixed, then Eq.~(\ref{10-3}) gives the lower
velocity $u_0$ below which -- and 
Eq.~(\ref{10-4}) gives the upper velocity above which -- the solution we
are interested in disappears. These critical values of $u_0$ are
functions of the incoming density $n_0$ and of 
the potential strength $\kappa$ determined by the equations
\begin{equation}\label{10-5}
    F\left(\frac{u_0}{\sqrt{n_0}}\right)=
\frac{\kappa}{\sqrt{n_0}}\quad\text{(lower boundary)},
\end{equation}
and
\begin{equation}\label{10-6}
    \frac{u_0}{\sqrt{n_0}}\,
F\left(2\frac{\sqrt{n_0}}{u_0}-1\right)=\frac{\kappa}{\sqrt{n_0}}
    \quad\text{(upper boundary)},
\end{equation}
where we recall that function $F$ is defined by (\ref{8-11}). These
two curves are plotted in Fig.~\ref{fig3}.

For small $\kappa/\sqrt{n_0}$ we get
the series expansions:
\begin{equation}\label{10-7}
    \frac{u_0}{\sqrt{n_0}}\simeq 1-
\frac32\left(\frac{\kappa}{\sqrt{n_0}}\right)^{2/3},
\quad\text{(lower boundary)},
\end{equation}
and
\begin{equation}\label{10-8}
    \frac{u_0}{\sqrt{n_0}}\simeq 1+
\frac34\left(\frac{\kappa}{\sqrt{n_0}}\right)^{2/3},
\quad\text{(upper boundary)}.
\end{equation}
In the opposite limit of large $\kappa/\sqrt{n_0}$ we get
\begin{equation}\label{11-1}
    \frac{u_0}{\sqrt{n_0}}
\simeq \frac1{2\sqrt{2}}\left(\frac{\kappa}{\sqrt{n_0}}\right)^{-2},
\quad\text{(lower boundary)},
\end{equation}
and
\begin{equation}\label{11-2}
    \frac{u_0}{\sqrt{n_0}}\simeq 2-
2\sqrt{2}\left(\frac{\kappa}{\sqrt{n_0}}\right)^{-2},
\quad\text{(upper boundary)}.
\end{equation}

We emphasize here that Eqs.~(\ref{10-5}) and (\ref{10-6}) are
important results of our study because they determine the region of
parameters for which our analytical solution exists, i.e.,
for which (i) the DSW is expelled
upstream from the obstacle and (ii) the flow forms in vicinity of the obstacle
a sonic horizon such as schematically represented
in Fig.~\ref{fig0}.

In this line, it is important to notice that the above approach is not
applicable for positive values of the soliton edge velocity $V_+>0$,
i.e., when the soliton edge of the DSW is attached to the obstacle.
This is different from what happens in the case of thick obstacles
\cite{legk} where the existence of a characteristic length $l$ of the
potential representing the obstacle plays a crucial role. Since in
this case $l$ is large compared with the wavelength of the (possibly
attached) DSW, the regions of DSW and of the ``hydraulic solution'' in
vicinity of the obstacle are well separated and one can safely assume
formation of a plateau at the left boundary of the region of the
hydraulic solution. In the present case of a $\delta$ potential such a
plateau forms at a distance of the same order of ma\-gni\-tu\-de as the DSW
wavelength and, hence, in the case where the DSW is attached to the
obstacle the solution of region $C$ (see Fig.~\ref{fig1}) completely
disappears and the formulae derived above loose their meaning.

\section{Comparison with the flow past a wide penetrable
  barrier}\label{secV}

\begin{figure}[bt]
\begin{center}
\includegraphics*[width=0.99\linewidth]{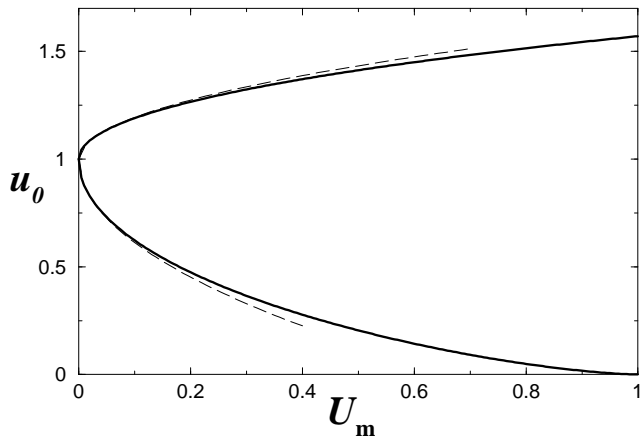}
\caption{Boundaries of the region (\ref{18-6}) for which the hydraulic
  solution exists and the upstream DSW is detached from a thick and
  smooth obstacle (here we use units such that $n_0=1$). The dashed
  lines correspond to the approximations (\ref{18-7}) which are valid
  when $U_{\rm m}\ll 1$.}
\label{fig6}\end{center}
\end{figure}

In this section we briefly compare the results we obtained for a thin potential
with those obtained in Ref. \cite{legk} for a
wide and smooth potential $U(x)$ which takes its maximal value at $x=0$,
\begin{equation}\label{17-2}
    U_{\rm m}=\mathrm{max}\{U(x)\}=U(0),
\end{equation}
and differs from zero only inside the region
\begin{equation}\label{17-3}
    -l\lesssim x\lesssim l.
\end{equation}
If $l$ and all lengths characteristic of the potential
are much larger than the healing length of the condensate, the
transition from an upstream subsonic flow to a downstream supersonic
one is described by the so-called ``trans-critical flow''
\cite{legk}. It is obtained for an incoming velocity $u_0\in[u_-,u_+]$,
where the critical velocities $u_{\pm}$ are the roots of the equation
\begin{equation}\label{18-2}
    \tfrac12u^2-\tfrac32u^{2/3}+1=U_{\rm m}
\end{equation}
(to simplify the notation, in this Section we assume
$n_{\rm ref}=n_0$). If $U_{\rm m}\ll1$, the critical velocities are given by the
series expansions
\begin{equation}\label{18-3}
    u_{\pm}\approx 1\pm\sqrt{\frac{3U_{\rm m}}2}.
\end{equation}
Flow velocities $u_{1,2}$ at the boundaries of the hydraulic region
are roots of the equation
\begin{equation}\label{18-4}
    \frac{u^2}2+\left(\frac{u_0-u}2+1\right)^2-
\frac32\left[u\left(\frac{u_0-u}2+1\right)^2
    \right]^{2/3}=U_{\rm m}.
\end{equation}
The smaller root corresponds to the upstream velocity $u_1$ and
the larger one to the downstream velocity $u_2$. If $U_{\rm m}\ll 1$
we get
\begin{equation}\label{18-5}
\begin{split}
    u_1=1+\frac13(u_0-1)-\sqrt{\frac{2U_{\rm m}}3},\\
    u_2=1+\frac13(u_0-1)+\sqrt{\frac{2U_{\rm m}}3}.\end{split}
\end{equation}
The DSWs can be attached to the trans-critical flow as accounted for in
detail in \cite{legk}.  In our initial value problem, with a wave packet
incoming from the left infinity, there can be no downstream
DSW. However, an upstream DSW exists. It is detached from the obstacle
only if the condition $u_0<2-u_1$ is fulfilled. Thus, the solution we
are interested in, with a DSW expelled upstream from the region
of the obstacle, corresponds to a region in the plane $(U_{\rm m},u_0)$ where
\begin{equation}\label{18-6}
    u_-<u_0<2-u_1 \; .
\end{equation}
This determines the parameters of the flow and of the potential
leading to the formation of an acoustic analog of a black hole in
the flow of a condensate past a thick obstacle. If $U_{\rm m}\ll1$,
then inequalities (\ref{18-6}) take the form
\begin{equation}\label{18-7}
    1-\sqrt{\frac{3U_{\rm m}}2}<u_0<1+\sqrt{\frac{3U_{\rm m}}8}.
\end{equation}
The whole region (\ref{18-6}) is shown in Fig.~\ref{fig6} which is the
thick obstacle analog of Fig.~\ref{fig3}.

\section{Comparison with numerical simulations}\label{secIV}

Our analytical theory is based on the assumption that the stationary
flow with a plateau in region $C$ is formed instantaneously after
collision of the condensate with the obstacle. This assumption
justifies the application of the self-similar solution [characterized
by formulae (\ref{6-1}) and (\ref{6-1a})] which only depends on the
variable $x/t$. However, in practice (i) any initial wave-packet
has---contrarily to our simplified initial flow (\ref{3-8}) ---a
finite region of transition from a constant plateau density to vacuum,
and (ii) the stationary solution in region $C$ forms in a period of
time negligibly small compared with the asymptotic values of time
under consideration, but nonetheless finite, and this may induce some
noticeable effects. Therefore we performed numerical simulations for
determining the limitations of our analytical approach.

In our numerics the incident condensate
wave packet was represented by the following initial density distribution:
\begin{equation}\label{15-2}
    n(x,t=0)=n_0\left[\frac{1-\tanh(x/\Delta)}{2}\right]^4 \; .
\end{equation}
According to the prescription of Sec. \ref{secII} we take here $n_{\rm
  ref}=n_0$, so that $n_0=1$ henceforth. The velocity of the incident
flow it taken to be $u_0=1$ (i.e., $u_0=c_0=c_{\rm ref}$). We consider three
different cases with different thicknesses $\Delta$ of the incident
front of the wave packet: $\Delta=20$, 10 and 5 (in units of $\xi_0=
\xi_{\rm ref}$).
The obstacle is modeled by a Gaussian potential
\begin{equation}\label{15-1}
    U(x)=U_0\, \exp\left\{-x^2/\sigma^2\right\}\,
\end{equation}
with $U_0=4$ and $\sigma=0.5$ (i.e., the size of the
obstacle is less than the healing length).  

\begin{figure}[bt]
\begin{center}
\includegraphics*[width=0.99\linewidth]{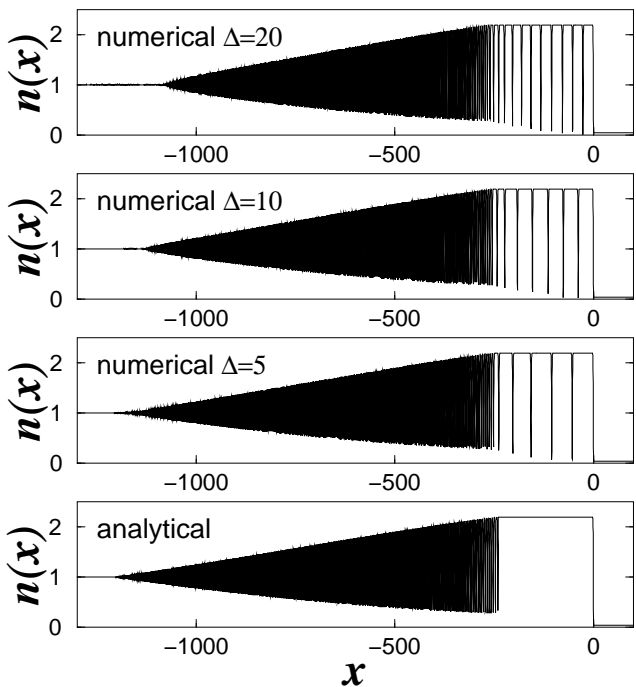}
\caption{Density profiles at $t=500$. The three upper panels present
  the results of numerical simulations performed in the case where the
  obstacle is represented by the Gaussian potential (\ref{15-1}) and
  the incident beam by (\ref{15-2}) with $\Delta=20$, $10$
  and $5$. The lower panel is the analytical result corresponding to a
  point-like obstacle (\ref{3-7}) and a step-like incident beam
  (\ref{3-51}),(\ref{3-52}), i.e., $\Delta=0$.}\label{fig4}
\end{center}
\end{figure}

\begin{figure}[bt]
\begin{center}
\includegraphics*[width=0.99\linewidth]{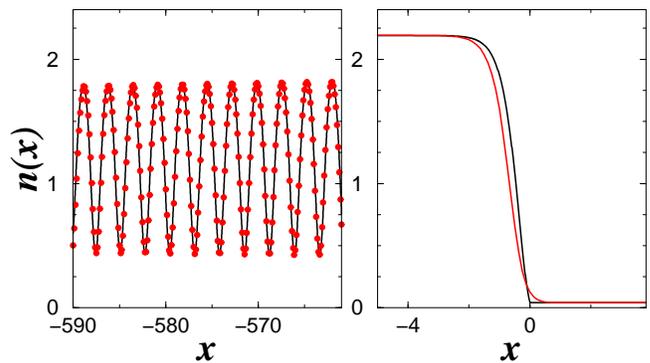}
\caption{(Color online) Density profiles at $t=500$. These plots
  are enlargements of specific parts of the two lower plots of
  Fig.~\ref{fig4}. The left plot concerns a part of the DSW
  ($-590<x<-560$), and the right one concentrates on the region of the
  obstacle. In both the right and left plots the black solid line
  corresponds to the analytical density profile. In the right plot
  the red solid line is the numerical density computed in the case
  $\Delta=5$. The numerical profile is represented by red points in the
  left plot, because it would be indistinguishable from the analytical
  result if represented with a solid line.}\label{fig4bis}
\end{center}
\end{figure}

The results of the numerical evolution of the flow after a time
$t=500$ are shown in Fig.~\ref{fig4}, where the different numerical
curves correspond to different values of $\Delta$. The potential
(\ref{15-1}) deviates significantly from a $\delta$ potential: we have
here $\sigma=0.5$ which is too large for using the natural
prescription for defining, starting from (\ref{15-1}), an equivalent
$\delta$ potential of the form (\ref{3-7}) by $\kappa= \int U(x){\rm
  d}x=U_0\sigma\sqrt{\pi}=3.5$. This prescription would be accurate
only in the limit $\sigma\ll 1$. The value $\kappa=5.2$ used in the
analytical procedure results from a fit of the numerically obtained
density of the plateau ($n_1=2.192$); then all the other parameters of
the flow are determined by the analytical formulae. That this
procedure is legitimate is confirmed by the very good description of
the upstream velocity $u_1$ in region $C$ and of the density $n_2$ and
velocity $u_2$ in the supersonic downstream region $D$: one obtains
analytically $u_1=0.0390$, $n_2=0.0412$ and $u_2=2.075$, whereas one
finds numerically $u_1=0.0388$, $n_2=0.0410$ and $u_2=2.075$ (see also
Fig.~\ref{fig4bis}). 

We checked that the natural prescription $\kappa=U_0\sigma\sqrt{\pi}$
yields a very good agreement of the analytical and of the numerical
approaches for low values of $\sigma$. For instance when $U_0=4$ and
$\sigma=0.1$ the natural prescription yields $\kappa=0.71$ and for
this value of $\kappa$ one obtains analytically $n_1=1.658$, whereas
one find numerically $n_1=1.6703$. In this case perfect agreement with
the numerical va\-lues of $n_1$, $n_2$, $u_1$ and $u_2$ is found by
using an effective $\kappa=0.7365$, close to the natural prescription.
However, for reducing the numerical effort we performed extensive
simulations only in the case $\sigma=0.5$. The efficiency of the
effective $\delta$ barrier for representing a potential of the form
(\ref{15-1}) even though $\sigma$ is not very small is a positive test
of the robustness of our analytical description of the system.

We note that the theory predicts the conservation of the Riemann
invariant
\begin{equation}\label{16-1}
\la_+^{\sss A}(=\la_0=1.5)=\la_+^{\sss C}=u_1/2+\sqrt{n_1}
\end{equation}
across the DSW (cf. Fig. \ref{fig2}) and that this is fulfilled with
very good accuracy in the numerical simulations. The positions
$X_{\pm}(t)$ of the edges of the DSW are also in good agreement with
the theoretical prediction.  Because of the scale used on the $x$-axis
of Fig.~\ref{fig4}, one could think that the density is discontinuous
around $x=0$. This is not the case, and furthermore one can verify
that in this region the analytical and numerical density profiles are
very similar, as shown in Fig.~\ref{fig4bis} (right plot). Also inside
the DSW, the nonlinear oscillations are very well described by the
analytical approach (same figure, left plot).

Hence, we can legitimately state that the hypothesis of rapid
formation of the DSW which, as already explained, is at the heart of
our assumption of self-similarity, is va\-li\-da\-ted by the
comparison with numerical simulations. This makes it possible to
bypass the delicate and in\-te\-res\-ting question of the short time
dynamics which is discussed for instance in
Ref. \cite{Dekel2010}. However, our numerical simulations reveal a
peculiar feature of the flow, namely, a train of dark solitons is
observed along the plateau in region $C$ (cf. Fig.~\ref{fig4},
upper panels).  As it is clear from the figure, the number of solitons
in the plateau decreases with $\Delta$ and the appearance of a train
of solitons is thus an effect of the finite width of the front of the
initial wave packet. These solitons are generated at the initial stage
of evolution and their precise characteristics depend on the details
of the initial density distribution. We note that the number of
solitons in the train is several orders of magnitude lower than the
number of nonlinear oscillations in the DSW (see Fig.~\ref{fig4}),
i.e., the existence of this train of solitons is a minor effect in our
asymptotic description of the flow. Nonetheless, in any experiment the
incident wave packet will not be infinitely sharp and the resulting
train of solitons should be easily observed. It is therefore worth
spending some time discussing its properties.

The periodic solution (\ref{4-2}), (\ref{4-3}) describes a train of
solitons if one considers a situation where $\la_2=\la_3$ while the
other Riemann invariants ($\la_1$ and $\la_4$) remain constant. Here
one has
\begin{equation}\label{16-2}
    \la_1=-\la_0\cdot\frac{2-M_{\s2 c}}{2+M_{\s2 c}},\qquad \la_4=\la_0\; .
\end{equation}
The velocity $V$ and minimal density $n_{\rm min}$ of one of the
solitons of the train can be expressed as functions of the coinciding
Riemann invariants $\la_2(x,t)=\la_3(x,t)$:
\begin{equation}\label{16-3a}
    V=\frac12(\la_1+2\la_3+\la_4)=\frac{u_1}2+\la_3,\end{equation}
and
\begin{equation}\label{16-3b}
\begin{split}
n_{\rm min}=&\frac12(\la_1+\la_4)^2-\la_3(\la_1+\la_4-\la_3)\\
=&\frac74u_1^2-u_1V+V^2,
    \end{split}
\end{equation}
where $u_1$ is given by Eq.~(\ref{9-6}), $M_{\s2 c}$ being determined
as a function of $\kappa$ and $\la_0$ by Eq.~(\ref{9-5}). In the
present case ($n_0=1$, $u_0=1$ and $\kappa=5.2$) one gets
$u_1=0.039$. 
In accordance with the general spirit of
our approach, we again assume self-similarity of the
solution, i.e., the locations of the solitons are given by an equation
of the form $x=Vt$. The corresponding values of
\begin{equation}\label{17-1}
    \la_2=\la_3=\frac{x}t-\frac{u_1}2
\end{equation}
are shown in Fig.~\ref{fig2} as a (red online) dashed line attached to
the soliton edge of the DSW \cite{note-1}. Note that (\ref{17-1})
corresponds to the solution (\ref{6-1a}) of the Whitham equation in
the soliton limit where $\la_2(x/t)=\la_3(x/t)$.

\begin{figure}[bt]
\begin{center}
\includegraphics*[width=0.99\linewidth]{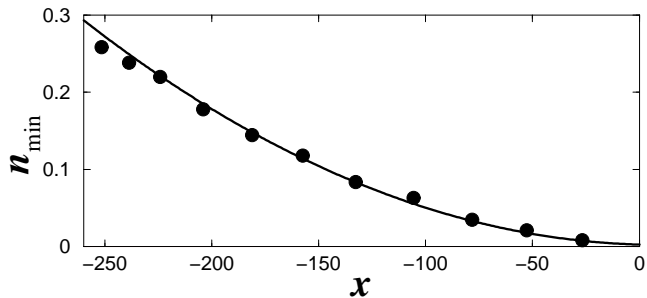}
\caption{The thick solid line represents the analytical prediction
  (\ref{16-4}) for the positions of the points of minimal density of
  the soliton train at time $t=500$. The value $u_1=0.039$ was used in
  the plot of Eq.~(\ref{16-4}) [see discussion below
  Eq.~(\ref{16-3b})].  The dots show the same quantity obtained from
  the density profile $n(x,t=500)$ computed numerically in the case
  $\Delta=20$ (see Fig.~\ref{fig4}, upper panel).}
\label{fig5}\end{center}
\end{figure}

Then, elimination of $V$ from (\ref{16-3b}) yields the
following relation
\begin{equation}\label{16-4}
    n_{\rm min}=\frac74u_1^2-u_1\cdot\frac{x}t+\left(\frac{x}t\right)^2
\end{equation}
between two easily measurable parameters: the minimal density $n_{\rm
  min}$ of a soliton and its coordinate $x$ at time $t$. 
The very good agreement of this analytical prediction with the
numerical results is illustrated in Fig.~\ref{fig5} where the solid
line represents the curve (\ref{16-4}) whereas the dots show the
minimal density at the position of the solitons as observed in
Fig.~\ref{fig4}, upper panel (in the case where $\Delta=20$).

Hence, although the detailed description of all the features of the
soliton train---depending of the precise shape of the incident front
of the wave-packet and of the potential---is beyond the scope of our
approach, we still can get some insight on the loci of the density
minima by assuming a kind of self similarity which relies on the fact
---confirmed by inspection of the numerical simulations---that the
train is formed in a very short period, at the initial stage of the
collision between the condensate and the obstacle. It then
evolves according to the solitonic limit of Whitham equations.

\section{Conclusions}\label{discussion}

In the present work we studied the flow of a Bose-Einstein beam
incident onto an obstacle in a regime where a structure forms similar
to the supersonic expansion obtained in a Laval nozzle. For an
appropriate choice of incident flow and of potential modeling the
obstacle we showed that, at large time, an acoustic horizon is created
around the obstacle. Its formation is accompanied by a release of
energy in the form of a dispersive shock wave whose characteristics
were studied in detail. This made it possible to precisely determine
the condition of formation of the structure studied in the present
work (see Fig.~\ref{fig3}).

The analytical description has been sustained by numerical simulations
which confirmed our analysis. The numerics however revealed an
interesting and unexpected feature: the ejection of the DSW from the
region of the horizon is typically accompanied by a train of dark
solitons which are only slowly expelled from the obstacle. We argued
that this train of solitons is formed in the initial stage of
collision of the Bose-Einstein beam with the obstacle and that its
detailed characteristics depend on the precise shape of the incident
beam and of the obstacle potential.  We also showed that one of its
important features can be described analytically (namely the line of
location of the density minima), precisely because this train is
formed in an early stage. 

The existence of this history-dependent train of solitons is due to
the specific dispersive and nonlinear wave mechanics describing the
Bose-Einstein condensate.  
It remains to investigate whether this train of solitons
does or does not hinders the experimental observation of
sonic analog of Hawking radiation.

\begin{acknowledgments}
  We would like to thank warmly I. Carusotto with whom we first
  discussed and observed numerically the formation of the dispersive
  shock structure presented in this work, who provided us with his
  numerical code and made judicious remarks on the
  manuscript. A.M.K. thanks LPTMS (Universit\'e Paris Sud and CNRS),
  where this work was done, for kind hospitality and financial
  support. This work was supported by the IFRAF Institute, by Grant
  ANR-08-BLAN-0165-01 and by RFBR (grant 09-02-00499). The research
  presented in this work was partly performed in the framework of the
  associated European laboratory ``Theoretical Physics and Condensed
  Matter'' (\sc{ens-landau}).
\end{acknowledgments}


\begin{thebibliography}{99}

\bibitem{landau} L. D. Landau, J. Phys. (USSR) {\bf 5}, 71 (1940);
  {\bf 11}, 91 (1947), reprinted in I. M. Khalatnikov, {\it An
    Introduction to the Theory of Superfluidity}, (Perseus Publishing,
  Cambridge, 2000).
\bibitem{Wilks} J. Wilks, {\it The properties of liquid and solid
    helium}, international series of monographs on physics, (Clarendon
  Press, Oxford, 1967).
\bibitem{feynman} R. P. Feynman, in {\it Progress in Low Temperature
    Physics,} edited by C. J. Gorter (North-Holland, Amsterdam, 1955),
  Vol. I, p.~17.

\bibitem{WMCA-99} T. Winiecki, J. F. McCann, and C. S. Adams, Phys. Rev. Lett.
{\bf 82}, 5186 (1999).

\bibitem{Car-2006} I. Carusotto, S. X. Hu, L. A. Collins, and A. Smerzi,
Phys. Rev. Lett. {\bf 97}, 260403 (2006).

\bibitem{GEGK-2007} Yu. G. Gladush, G. A. El, A. Gammal, and A. M. Kamchatnov,
Phys. Rev. A {\bf 75}, 033619 (2007).

\bibitem{GSK-2008} Yu. G. Gladush, L. A. Smirnov, and A. M. Kamchatnov,
J. Phys. B {\bf 41}, 165301 (2008).

\bibitem{Raman-1999} C. Raman {\it et al.}, Phys. Rev. Lett. {\bf 83},
  2502 (1999).

\bibitem{Amo-2009} A. Amo {\it et al.}, Nat. Phys. {\bf 5}, 805 (2009).

\bibitem{FPR-92} T. Frisch, Y. Pomeau, and S. Rica, Phys. Rev. Lett.
  {\bf 69}, 1644 (1992).

\bibitem{Inou-2001} S. Inouye {\it et al.}, Phys. Rev. Lett. {\bf 87}, 
080402 (2001).

\bibitem{Nee-2010}T. W. Neely, E. C. Samson, A. S. Bradley, M. J. Davis, 
and B. P. Anderson, Phys. Rev. Lett. {\bf 104}, 160401 (2010).

\bibitem{Nar-2011} G. Nardin {\it et al.}, Nat. Phys. {\bf 7}, 635 (2011).

\bibitem{San-2011} D. Sanvitto {\it et al.}, Nat. Phot. {\bf 5}, 610 (2011).

\bibitem{EGK-2006} G. A. El, A. Gammal, and A. M. Kamchatnov, Phys. Rev. Lett.
{\bf 97}, 180405 (2006).

\bibitem{kp-2008} A. M. Kamchatnov and L. P. Pitaevskii,
Phys. Rev. Lett. {\bf 100}, 160402 (2008).

\bibitem{kk-2011} A. M. Kamchatnov and S. V. Korneev, Phys. Lett. A
  {\bf 375}, 2577 (2011).

\bibitem{amo-2011} A. Amo {\it et al.}, Science {\bf 332}, 1167 (2011).

\bibitem{gro11} G. Grosso, G. Nardin, F. Morier-Genoud, Y. L\'eger,
  and B. Deveaud-Pl\'edran, Phys. Rev. Lett. {\bf 107}, 245301 (2011).

\bibitem{ek-2006} G.A. El and A.M. Kamchatnov, Phys. Lett.  A {\bf
    350}, 192 (2006); {\bf 352}, 554(E) (2006).

\bibitem{el-2009} G. A. El, A. M. Kamchatnov, V. V. Khodorovskii,
  E. S. Annibale, and A. Gammal, Phys. Rev. E {\bf 80}, 046317 (2009).

\bibitem{hi-2009} M. A. Hoefer and B. Ilan, Phys. Rev. A {\bf 80},
  061601(R) (2009).

\bibitem{Bur-1999} S. Burger {\it et al.} Phys. Rev. Lett. {\bf 83},
  5198 (1999).
\bibitem{Den-2000} J. Denschlag {\it et al}, Science {\bf 287}, 97
  (2000).
\bibitem{Ono-2000} R. Onofrio {\it et al.}, Phys. Rev. Lett. {\bf 85},
  2228 (2000).
\bibitem{ea-2007} P. Engels and C. Atherton, Phys. Rev. Lett. {\bf
    99}, 160405 (2007).
\bibitem{Dri-2010} D. Dries, S. E. Pollack, J. M. Hitchcock, and R. G. Hulet,
Phys. Rev. A {\bf 82}, 033603 (2010).
 
\bibitem{GACZ-2000} L. J. Garay, J. R. Anglin, J. I. Cirac, and P. Zoller,
Phys. Rev. Lett. {\bf 85}, 4643 (2000).

\bibitem{Gio-2004}S. Giovanazzi, C. Farrell, T. Kiss, and U. Leonhardt,
  Phys. Rev. A {\bf 70}, 063602 (2004).

\bibitem{BCGJ06} C. Barcelo, A. Cano, L.J. Garay, and G. Jannes, 
Phys. Rev. D {\bf 74}, 024008 (2006).

\bibitem{correlations} R. Balbinot, A. Fabbri, S. Fagnocchi,
  A. Recati and I. Carusotto, Phys. Rev. A {\bf 78}, 021603 (2008);
  I. Carusotto, S. Fagnocchi, A. Recati, R. Balbinot and A. Fabbri,
  New J. Phys. {\bf 10}, 103001 (2008).


\bibitem{RPC-2009} A. Recati, N. Pavloff, and I. Carusotto,
  Phys. Rev. A {\bf 80}, 043603 (2009).
\bibitem{Mac09} J. Macher and R. Parentani, Phys. Rev. A {\bf 80},
  043601 (2009).
\bibitem{Lar-2011} P.-\'E. Larr\'e, A. Recati, I. Carusotto, and N. Pavloff, 
Phys. Rev. A {\bf 85}, 013621 (2012).

\bibitem{lp-2001} P. Leboeuf and N. Pavloff, Phys. Rev. A {\bf 64},
  033602 (2001).

\bibitem{Laval} For a wide barrier the situation is analogous to the flow of
  a compressible gas through a Laval nozzle.

\bibitem{hakim} V. Hakim, Phys. Rev. E {\bf 55}, 2835 (1997).

\bibitem{pavloff} N. Pavloff, Phys. Rev. A {\bf 66}, 013610 (2002).

\bibitem{radouani} A. Radouani, Phys. Rev. A {\bf 70}, 013602 (2004).

\bibitem{legk} A. M. Leszczyszyn, G. A. El, Yu. G. Gladush, and
  A. M. Kamchatnov, Phys. Rev. A {\bf 79}, 063608 (2009).

\bibitem{LL-6} L. D. Landau and E. M. Lifshitz, {\it Fluid Mechanics}
  (Pergamon, Oxford, 1987).

\bibitem{Dekel2010} G. Dekel, V. Farberovich, V. Fleurov and
  A. Soffer, Phys. Rev. A {\bf 81}, 063638 (2010).

\bibitem{Men02} C. Menotti and S. Stringari, Phys. Rev. A {\bf 66},
  043610 (2002).

\bibitem{Hoe2006} M. A. Hoefer {\it et al.}, Phys. Rev. A {\bf 74},
  023623 (2006).

\bibitem{Mepp2009} R. Meppelink {\it et al.}, Phys. Rev. A {\bf 80},
  043606 (2009).


\bibitem{kamch2000}  A. M. Kamchatnov, {\it Nonlinear Periodic Waves and
Their Modulations---An Introductory Course} (World Scientific,
Singapore, 2000).

\bibitem{whitham74} G. B. Whitham,
{\it Linear and Nonlinear Waves} (Wiley--Interscience, New York, 1974).


\bibitem{fl86} M. G. Forest and J. E. Lee, Geometry and modulation theory for
periodic nonlinear Schr\"odinger equation, in {\it Oscillation Theory,
Computation, and Methods of Compensated Compactness,} Eds. C. Dafermos et al,
IMA Volumes on Mathematics and its Applications {\bf 2}, (Springer, N.Y., 1987).

\bibitem{pavlov87} M. V. Pavlov, Teor. Mat. Fiz. {\bf 71}, 351 (1987) [Theoret.
Math. Phys. {\bf 71}, 584 (1987)].

\bibitem{gke92}
A. V. Gurevich, A. L. Krylov and G. A. El,
{ Sov. Phys. JETP} {\bf 74}, 957  (1992).

\bibitem{gp74} A. V. Gurevich and L. P. Pitaevskii,
{Sov. Phys. JETP,} {\bf 38}, 291 (1974).

\bibitem{gs-1986} R. H. J. Grimshaw and N. Smyth, J. Fluid Mech. {\bf
    169}, 429 (1986).

\bibitem{smyth-1987} N. Smyth, Proc. Roy. Soc. Lond. Ser. A {\bf 409},
  79 (1987).

\bibitem{El-saeDSW} G. A. El, Chaos {\bf 15}, 037103 (2005); {\bf 16}, 029901 
(2006).

\bibitem{note-1} G. A. El kindly informed us that a similar effect has
  been observed in the numerical study of the evolution of a DSW
  propagating along a non-uniform background (unpublished).


\end{thebibliography}
\end{document}